\documentclass{appolb}
\usepackage{graphicx}
\usepackage{xcolor}
\usepackage{amsmath}
\usepackage{amssymb}
\usepackage{cite}
\usepackage{caption}
\usepackage[abs]{overpic}
\newcommand{\Pom}{\mathbb{P}}
\newcommand{\Reg}{\mathbb{R}}

\begin{document}
\title{Multiple pion pair production in a Regge-based model%
  \thanks{Presented at "Diffraction and Low-$x$ 2022", Corigliano Calabro
    (Italy), September 24-30, 2022.}%
}
\author{Rainer Schicker
\vspace{-.6mm}\address{Physikalisches Institut, University Heidelberg, Heidelberg}
\\[3mm]
  {in coll. with Laszlo Jenkovszky % of different affiliation
    
\vspace{-.6mm}\address{Bogolyubov ITP, National Academy of Sciences of Ukraine, Kiev}
}
}
\maketitle
\begin{abstract}
  Central diffractive event topologies at the LHC energies can be identified by
  two different approaches. First, the forward scattered protons can be measured
  in Roman pots. Second, a veto on hadronic activity away from midrapidity
  can be imposed to define a double-gap topology. Such a double-gap topology
  trigger has been implemented by the ALICE collaboration in Run 1 and
  Run 2 of the LHC. The analysis of these events allows to determine the
  charged-particle multiplicity within the acceptance. The excellent particle
  identification capabilities of ALICE allows to study two-track events both in
  the pion and kaon sector. Events with measured charged particle multiplicity
  larger than two can arise from multiple pair production. A Regge-based
  approach for modeling such multiple pair production is presented.

\end{abstract}
  
\section{Introduction}

Double-Pomeron fusion at hadron colliders results in a double-gap event
topology. Such a topology is defined by hadronic activity at or close to
midrapidity, and the absence thereof away from midrapidity. The multiplicity
distribution of such double-gap events has been measured in the ALICE central
barrel. To better understand such multiplicity distributions we present
here a Regge-based approach for multiple pion pair production in double-Pomeron
events. This model is based on a Dual Amplitude with Mandelstam Analyticity
(DAMA) \cite{DAMA}.  In this approach, the production of multiple pairs can
be modeled by including a Pomeron-Pomeron-Reggeon and a triple-Pomeron
coupling. The amplitude at Pomeron level within his DAMA formulation is given,
and the resulting mass distributions for double pion and double b-resonance
production are shown.

  \section{Multiplicity distribution of double-gap events}

  The charged-particle multiplicity in the ALICE central barrel has been
  analyzed in LHC Run 1 for both minimum bias and double-gap events
  \cite{Felix}.  
\begin{figure}[htb]
\centerline{%
\includegraphics[width=5.6cm]{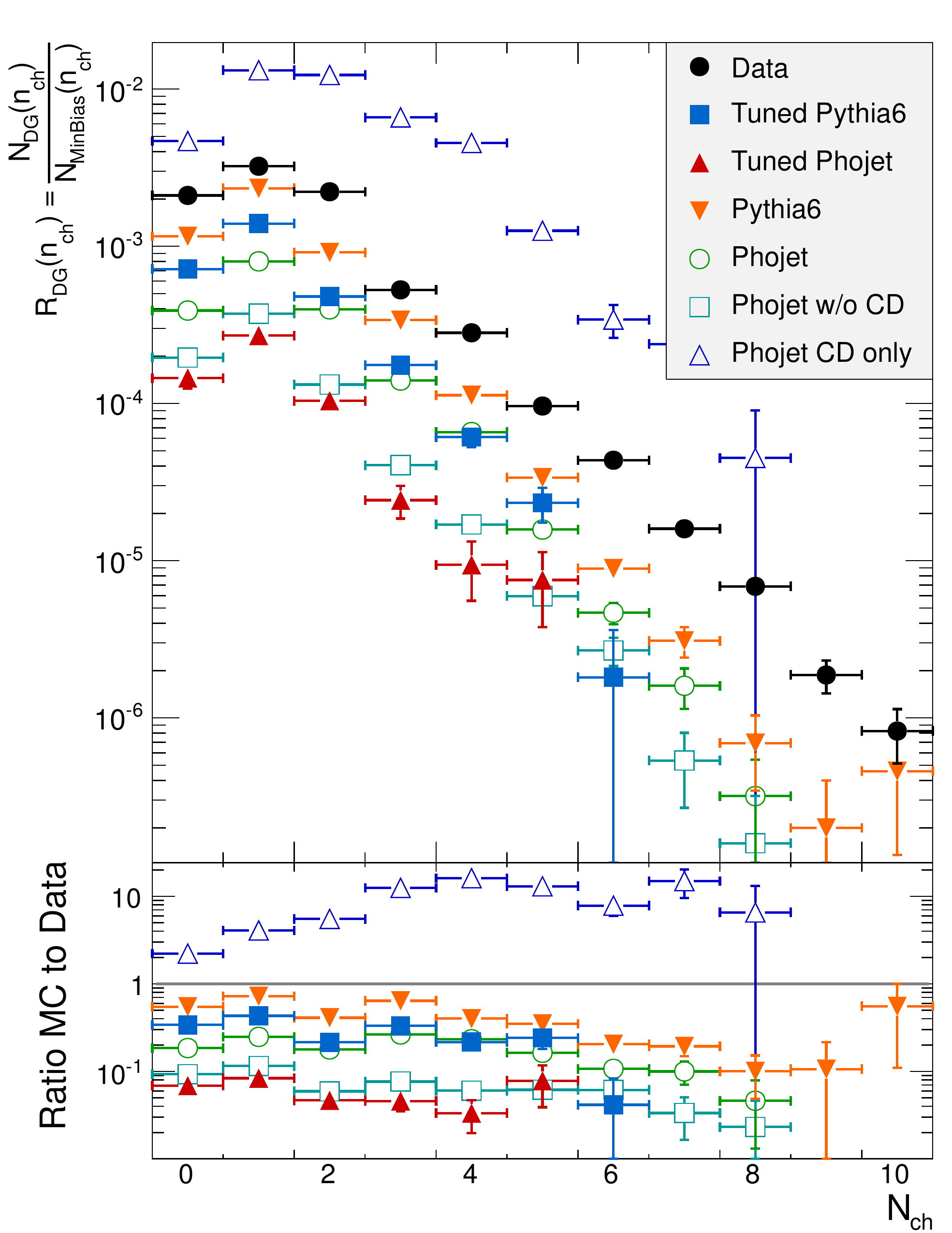}}
\caption{Double-gap probability in ALICE central barrel as function
  of charged-particle multiplicity (Figure taken from Ref. \cite{Felix}).}
\label{Fig:DG_frac}
\end{figure}
 
In Fig. \ref{Fig:DG_frac}, the probability of being a double-gap event is
shown as function of the charged-particle multiplicity N$_{ch}$ in the ALICE
central barrel. The ALICE data are shown in black circles, whereas the results
from Monte Carlo generators are shown in different colors. These probabilities 
clearly show a maximum at N$_{ch}$=1 and N$_{ch}$=2, demonstrating that
double-Pomeron events are dominated by very low multiplicities as compared to
minimum bias events. As indicated in this figure, none of the tested generators
shows reasonable agreement with the data. This discrepancy between the ALICE
measured double-gap events and the prediction of the tested generators motivates
the development of a model which can be used to analyze unlike-sign two-track
events resulting from single resonance decays, as well as the
higher-multiplicity events stemming from the decays of multiple resonances.

\section{A Regge model for double-Pomeron events}

The model for Pomeron-Pomeron-induced events presented in the following is
based on the DAMA approach. Pomeron-induced single-resonance production has
been presented in our previous studies \cite{FJS1,FJS2}. Here, we extend
this DAMA approach to the production of multiple resonances.

\vspace{.2cm}
\begin{minipage}[t]{.42\textwidth}
    \begin{overpic}[width=.9\textwidth]{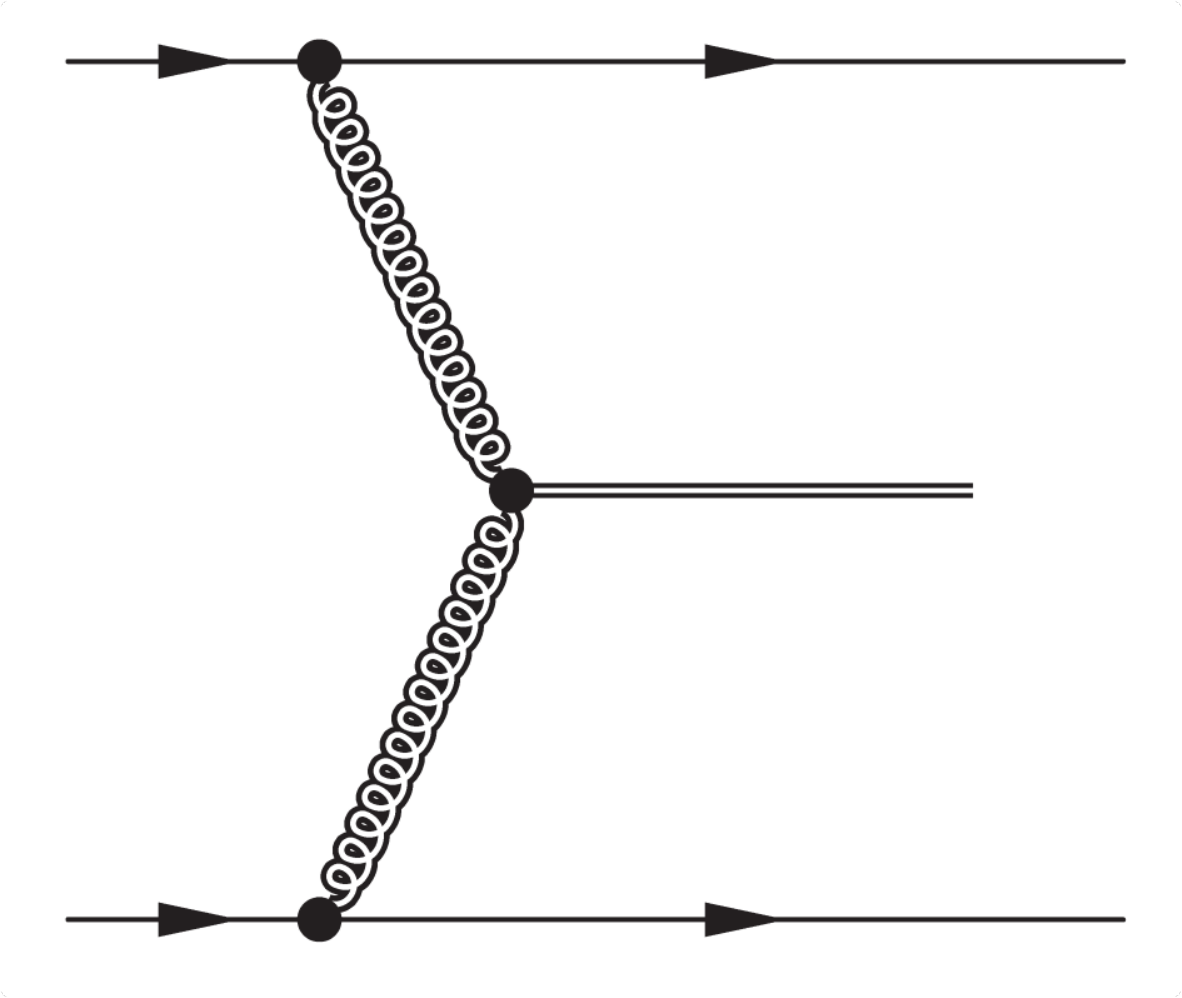}
    \end{overpic}
\end{minipage}
\thicklines
\begin{minipage}[t]{.16\textwidth}
    \begin{overpic}[width=.24\textwidth]{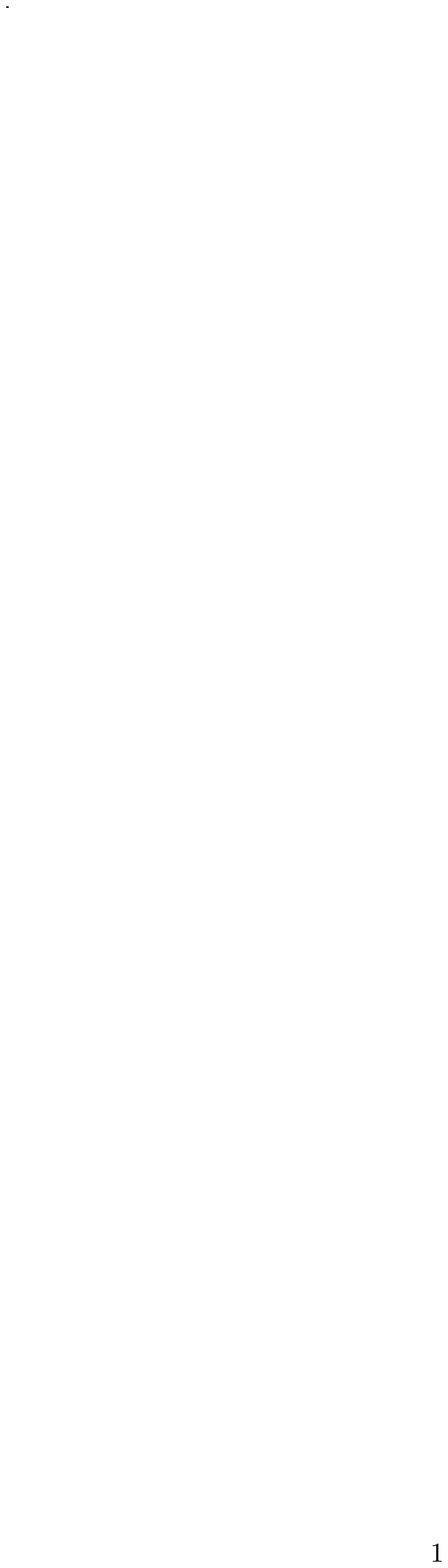}
    \put(6,20){\linethickness{.6mm}\vector(1,0){14.}}
      \put(5.,13.){amplitude}
      \put(4.5,8.8){subdiagram}
    \end{overpic}
 \end{minipage}   
\begin{minipage}[t]{.32\textwidth}    
  \hspace{0.2cm}
    \begin{overpic}[width=.9\textwidth]{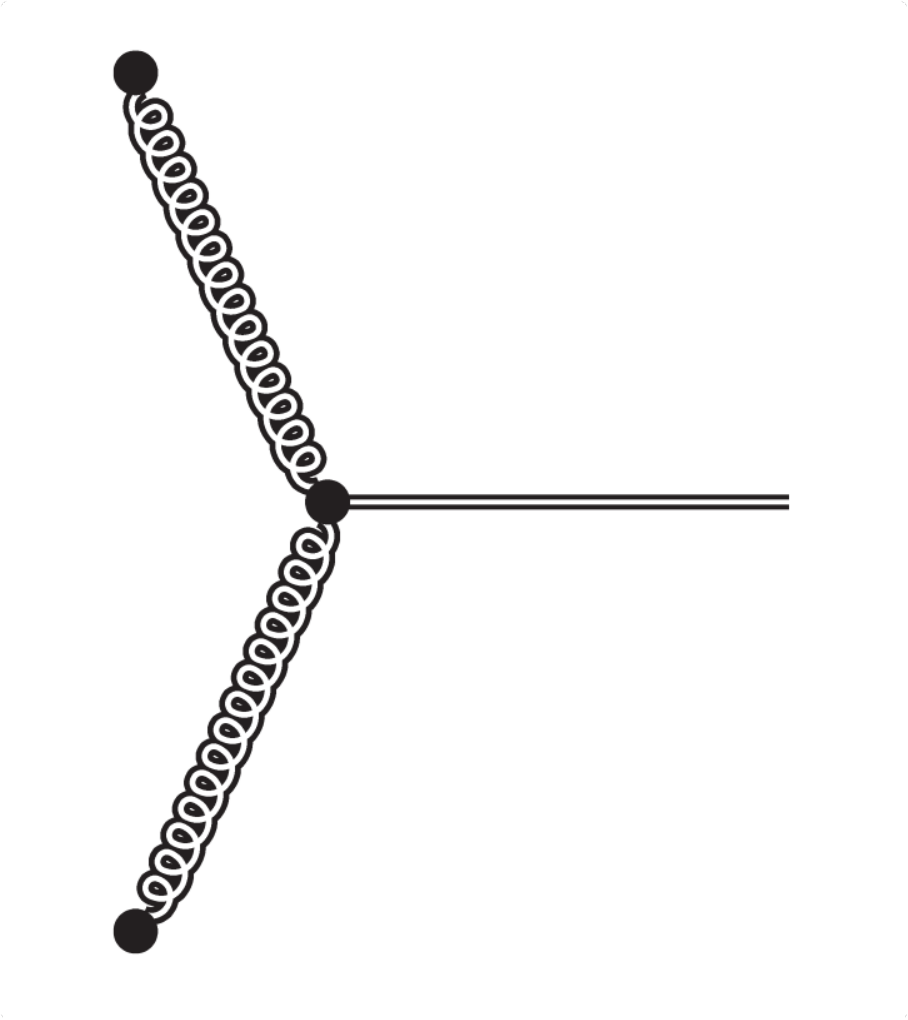}
    \end{overpic}
\end{minipage}
\vspace{-.2cm}
\begin{figure}[htb]
\caption{Amplitude at hadron level (left), and Pomeron subdiagram (right).}
\label{Fig:single_red}
\end{figure}

In Fig. \ref{Fig:single_red}, the amplitude for Pomeron-induced single-resonance
production at hadron level is shown on the left. The subdiagram on the right 
represents the amplitude for Pomeron-Pomeron $\rightarrow$ resonance. The cross 
section at hadron level is derived by convoluting the subdiagram cross section
with the Pomeron flux of the proton $F^{\Pom}_{\text{prot}}(t,\xi)$ defined by

\begin{equation}
F^{\Pom}_{\text{prot}}(t,\xi)=\frac{9\beta^{2}_{0}}{4\pi^{2}}[F_{1}(t)]^{2} \xi^{1-2\alpha(t)},
\label{Eq.1}
\end{equation}
with $F_{1}(t)$ the elastic form factor, and $\alpha(t)$ the Pomeron
trajectory \cite{FJS1}.
  
In the DAMA approach, multiple-resonance production can be modeled by
introducing a Pomeron-Pomeron-Reggeon ($\Pom\Pom\Reg$) coupling with subsequent
splitting of the intermediate Reggeon into the two final-state Reggeons.
Alternatively, the same final state can be formed by a triple-Pomeron
($\Pom\Pom\Pom$) coupling with the intermediate Pomeron decaying into
the two Reggeons.

%uncomment the following lines to place a figure

\vspace{.2cm}
\begin{minipage}[t]{.06\textwidth}
%  \begin{figure}
    \begin{overpic}[width=.08\textwidth]{empty.pdf}
    \end{overpic}
%  \end{figure}
\end{minipage}
 \begin{minipage}[t]{.44\textwidth}
%      \begin{figure}
%\vspace{-.3cm}
        %\begin{overpic}[grid,tics=4,width=.9\textwidth]{triple_reg_red.pdf}
          \begin{overpic}[width=.94\textwidth]{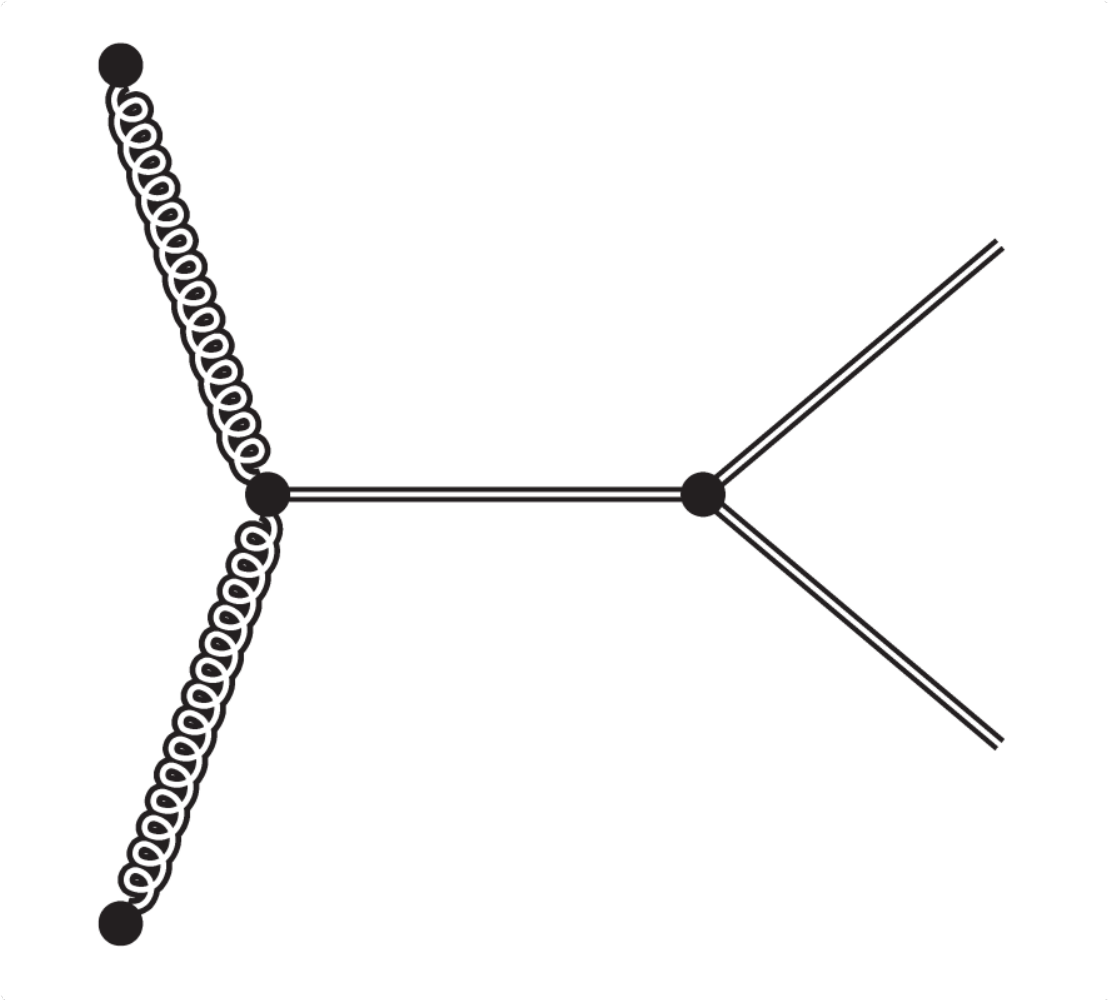}
            \put(3.,30.){$\alpha_{\Pom}$}
            \put(10.,37.){t$_{1}$}
     \put(2.0,10.0){$\alpha_{\Pom}$}
     \put(10.0,6.){t$_{2}$}
     \put(6.,23.0){g$_{1}$}
     \put(36.,23.0){g$_{2}$}
     %\linethickness{0.4mm}
      \put(-5.,23.6){\linethickness{.6mm}\vector(1,0){8}}
     %\put(-6.,18.6){\textcolor{black}{\vector(1,0){8}}}
     \put(-7.2,22.8){$\tilde{s}$}
     \put(18.,26.){$\alpha_{\Reg}(\tilde{s})$}
     \put(22.8,37.6){\linethickness{.6mm}\vector(0,-1){8}}
     \put(22.0,38.8){$\tilde{t}$}
     %\put(44.,20.0){{\Huge +}}
     \put(49.,21.6){{\Huge +}}
     \put(38.,38.){$\tilde{S}_{1}(M^{2}_{1})$}
     \put(37.,7.){$\tilde{S}_{2}(M^{2}_{2})$}
     \put(-14.,18.0){\large{($\tilde{s}\!=\!t_{1}\!+\!t_{2}$)}}
          \end{overpic}
%\end{figure}
     \end{minipage}
\begin{minipage}[t]{.44\textwidth}
%    \begin{figure}
   %   \vspace{-.4cm}
    \begin{overpic}[width=.94\textwidth]{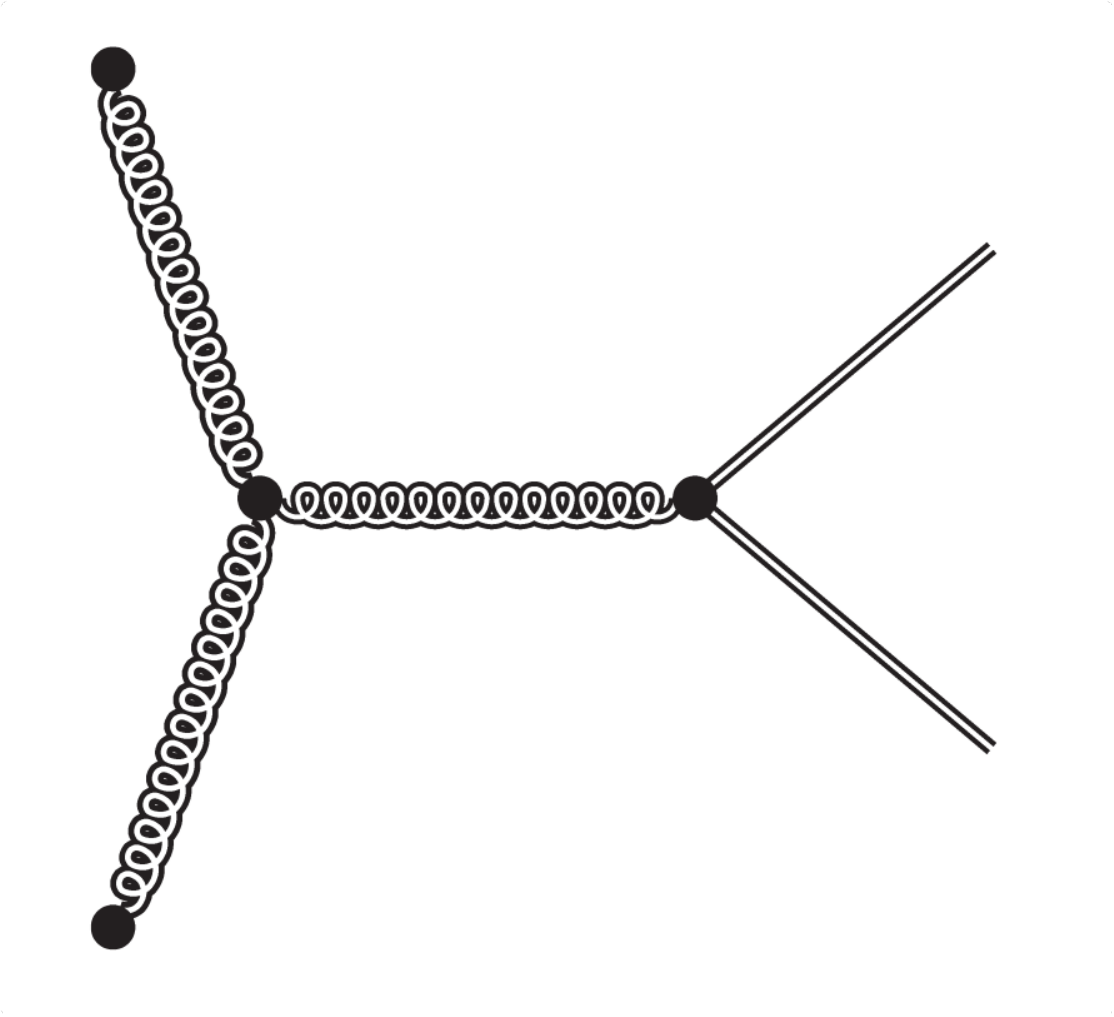}
     \put(6.,23.4){g$_{3}$}
     \put(36.,23.4){g$_{4}$}
      \put(19.,27.){$\alpha_{\Pom}(\tilde{s})$}
     \put(38.,38.){$\tilde{S}_{1}(M^{2}_{1})$}
     \put(38.,7.){$\tilde{S}_{2}(M^{2}_{2})$}
    \end{overpic}
%  \end{figure}
\end{minipage}
\vspace{-.4cm}
\begin{figure}[htb]
  \caption{Subdiagram for $\Pom\Pom\Reg$ amplitude (left), and $\Pom\Pom\Pom$ amplitude (right).}
\label{Fig:mult_red}
\end{figure}

The DAMA amplitude for the subdiagram shown in
Fig. \ref{Fig:mult_red} is given by

\begin{equation}
A_{\Pom\Pom\rightarrow \tilde{S}_{1}\tilde{S}_{2}}(\tilde{s},\tilde{t},M^{2}_{1},M^{2}_{2})) =
    \frac{1}{\sqrt{M^{2}_{1}M^{2}_{2}}} \sum_{\Pom\Pom\Reg,\Pom\Pom\Pom}\sum_{n}
    \frac{g_{i}g_{j}e^{b \alpha(\tilde{t})}}{n-\alpha(\tilde{s})},
\end{equation}
with the first summation over the two amplitudes of Fig. \ref{Fig:mult_red}
defined by the $\Pom\Pom\Reg$ coupling with g$_{i}$,g$_{j}$=g$_{1}$,g$_{2}$,  and
the $\Pom\Pom\Pom$ coupling with g$_{i}$,g$_{j}$=g$_{3}$,g$_{4}$.
The index $n$ sums over the spins of the resonances of the
intermediate trajectory which connects the vertices $i,j$. From this amplitude,
the cross section at Pomeron level is derived by the optical theorem
\begin{equation}
  \sigma_{t}(\tilde{s},M^{2}_{1},M^{2}_{2}) = \Im m\;
  A (\tilde{s},\tilde{t}\!=\!0,M^{2}_{1},M^{2}_{2}), 
\end{equation}
with the imaginary part of $A(\tilde{s},\tilde{t},M^{2}_{1},M^{2}_{2})$ defined by
$\alpha_{\Reg}(\tilde{s})$ and $\alpha_{\Pom}(\tilde{s})$ for the $\Pom\Pom\Reg$
and the $\Pom\Pom\Pom$ diagrams of Fig. \ref{Fig:mult_red}, respectively.

\section{Reggeizing $q\bar{q}$ states in the light quark sector}

The final-state mesons derive from the decay of the meson resonances lying on
the two Regge trajectories $\tilde{S}_1$ and $\tilde{S}_2$ as illustrated in 
Fig. \ref{Fig:mult_red}. In order to be able to include mesonic bound
states of different radial and orbital excitations, a unified description of
$q\bar{q}$ bound states in the different flavour sectors is needed. Such a
unified description of $q\bar{q}$ bound states including a confinement
potential, a spin-orbit, a hyperfine and an annihilation interaction is
presented in Ref. \cite{Isgur}. The solutions for these $q\bar{q}$ bound states
are given in spectroscopic notation $n\:^{2S+1}L_{J}$.
 
\vspace{4.mm}
   \begin{minipage}[h]{.34\textwidth}
    {\small spectr. notation $n\:^{2S+1}L_{J}$:} \newline
    {\small - $n$ radial quantum number} \newline
    {\small - $S$ spin} \newline
    {\small - $L$ orbital ang. momentum} \newline
    {\small - $J$ total ang. momentum}  
   \end{minipage}
   \hspace{1.mm}
      \begin{minipage}[t]{.62\textwidth}
%\begin{table}[h]
\begin{tabular}{|p{0.16\textwidth}|p{0.16\textwidth}|p{0.12\textwidth}|p{0.12\textwidth}|p{0.12\textwidth}|}
\hline
\small{$n\:^{2S+1}L_{J}$} & \small{mass} & \small{PDG} & \small{mass} & \small{width}  \\
 & \small{Ref. \cite{Isgur}} & & \small{(PDG)} & \small{(PDG)}  \\
\hline
\small{$1^{1}S_{0}$}& \small{150} & \small{$\pi$} & \small{140} & \small{0}  \\
\small{$1^{1}P_{1}$}& \small{1220} & \small{$b_{1}$} & \small{1230} & \small{142}  \\
\small{$1^{1}D_{2}$}& \small{1680} & \small{$\pi_{2}$} & \small{1672} & \small{258}  \\
\small{$1^{1}F_{3}$}& \small{2030} & \small{------} & \small{------} & \small{------}  \\
\small{$1^{1}G_{4}$}& \small{2330} & \small{------} & \small{------} & \small{------}  \\
\hline
\end{tabular}   
%\end{table}
      \end{minipage}
%   \begin{minipage}[h]{.24\textwidth}
    % \begin{overpic}[width=.14\textwidth]{empty.pdf}
    %   \end{overpic}
%   \end{minipage}
      \begin{table}[h]
%\centering\captionsetup{justification=RaggedLeft}
\caption{Masses and widths in MeV.}
\label{table:t1}
      \end{table}
 
 In Table \ref{table:t1}, masses are presented for the isovector channel in the
 light quark sector for the radial ground state for S,P,D,F and G-wave, and are
 compared to the values given by the Particle Data Group \cite{PDG}.
 The S- and D-wave bound states calculated in Ref. \cite{Isgur} are identified
 with the  $\pi$ and the $\pi_{2}$ states of mass 140 and 1672 MeV,
 respectively. The P-wave solution is associated to the  known $b_{1}$ state
 of mass 1230 MeV. No candidates for the predicted F- and G-wave bound
 states have so far been experimentally identified \cite{PDG}.
 
\section{Non-linear complex Regge trajectory}

The small but existing non-linear dependence of the \mbox{spin of a resonance}
to its mass squared can be used to make a Regge trajectory $\alpha(M^{2})$
\mbox{a complex} entity with real and imaginary parts being related by a
dispersion \mbox{relation \cite{Disp}.} Here, the real part is defined by the
value of the spin, and the imaginary part is related to the decay
width $\Gamma$ by
$\Im m\: \alpha(M_{R}^{2}) = \Gamma(M_{R})\:\alpha^{'}\:M_{R} $, with
$\alpha^{'}$ denoting the derivative of the real part of the trajectory. In a
simple model, the imaginary part is chosen as a sum of single threshold terms

 \begin{equation}
\Im m\: \alpha(s)\!=\!\sum_{n}\!c_{n} (s\!-\!s_{n})^{1/2} 
\big(\frac{s\!-\!s_{n}}{s}\big)^{\!|\Re e\:\alpha(s_{n})|} \theta(s\!-\!s_{n}).
\label{eq:imag}
\end{equation}

 In Eq. \ref{eq:imag}, the coefficients $c_{n}$ are fit parameters, and
 the parameters $s_{n}$ represent kinematical thresholds of decay channels. 
 
 \subsection{The ($\pi,b$)-trajectory}

 A Regge trajectory, called the ($\pi,b$)-trajectory hereafter, is defined
 by the values of mass and width of the S, P and D-waves shown in
 Table \ref{table:t1}.
 
  \begin{minipage}[t]{.98\textwidth}
%    \begin{center}
   %  \begin{figure}
    \vspace{.1mm}
    \begin{overpic}[width=.48\textwidth]{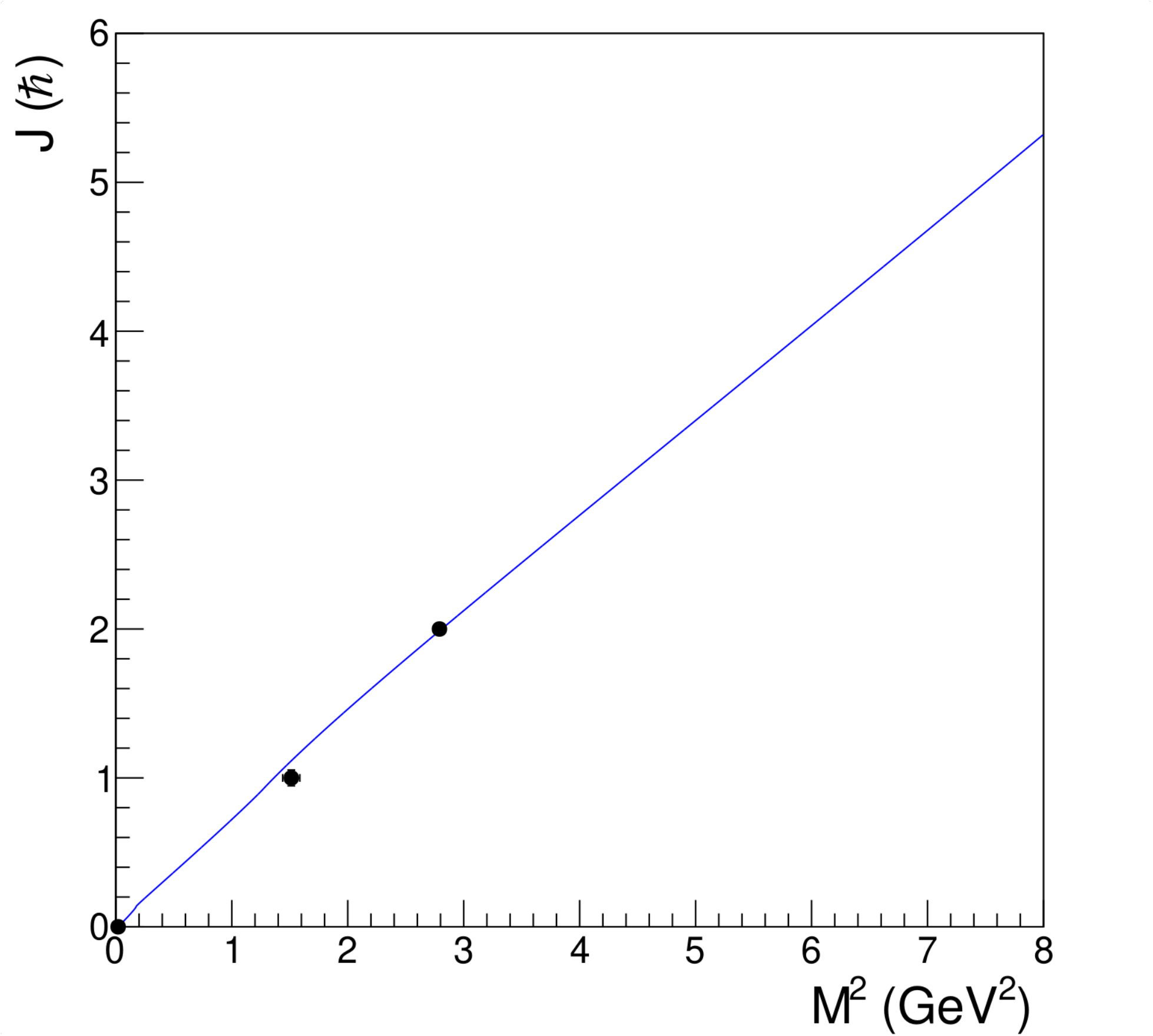}
%\put(20.0,8.){\footnotesize real part $(\pi,b)$-traj.}
\multiput(6.0,27.8)(1.6,0){16}{\textcolor{red}{-}}
\multiput(6.0,35.45)(1.6,0){22}{\textcolor{red}{-}}
\multiput(6.0,43.1)(1.6,0){28}{\textcolor{red}{-}}
%\multiput(4.8,23.75)(1.6,0){14}{\color{red}\linethickness{0.2mm}\line(1,0){.8}}
%\multiput(4.8,30.0)(1.6,0){19}{\color{red}\linethickness{0.2mm}\line(1,0){.8}}
%\multiput(4.8,36.4)(1.6,0){24}{\color{red}\linethickness{0.2mm}\line(1,0){.8}}
\put(31.8,28.6){\textcolor{red}{\circle{1.2}}}
\put(41.2,36.2){\textcolor{red}{\circle{1.2}}}
\put(50.7,43.9){\textcolor{red}{\circle{1.2}}}
\put(8.2,6.5){\textcolor{blue}{\footnotesize{$\pi$}}}
\put(15.9,12.4){\textcolor{blue}{\footnotesize{$b_{1}$}}}
\put(23.8,20.2){\textcolor{blue}{\footnotesize{$\pi_{2}$}}}
\put(33.2,27.4){\textcolor{red}{\footnotesize{$b_{3}$}}}
\put(42.2,35.3){\textcolor{red}{\footnotesize{$\pi_{4}$}}}
\put(50.8,40.8){\textcolor{red}{\footnotesize{$b_{5}$}}}
    \end{overpic}
\hspace{.2cm} 
\begin{overpic}[width=.48\textwidth]{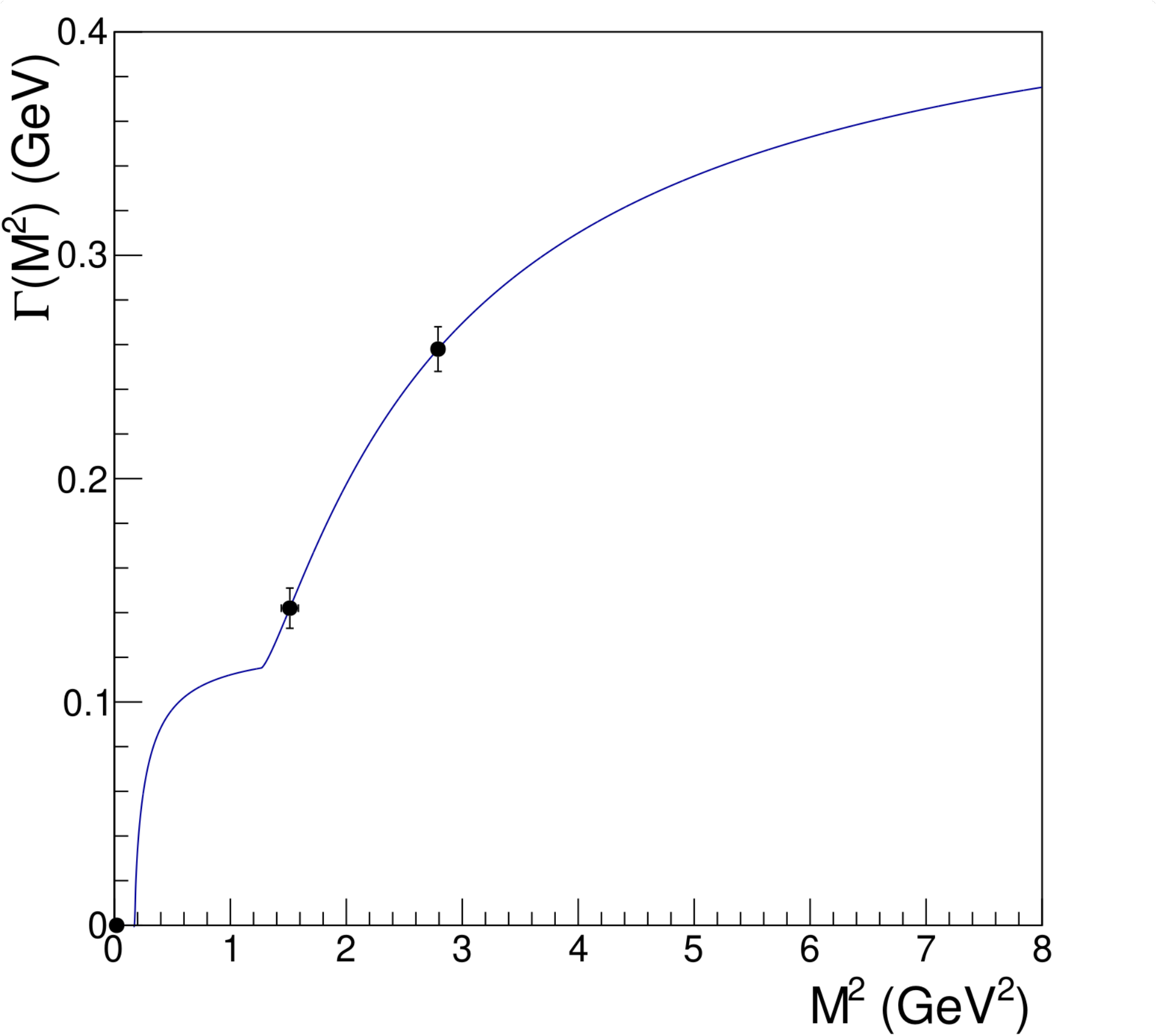}
%  \put(12.,8.0){\footnotesize imaginary part $(\pi,\!b)$-traj.}
%\put(20.,-7.4){\small $b_{1}\rightarrow K\bar{K}\pi:\;s_{1}\!=\!1.27\: \text{GeV}^{2}$}
%\put(16.,-12.4){\small $\pi_{2}\rightarrow 3\pi:\;s_{0}\!=\!0.176\: \text{GeV}^{2}$}
\put(4.2,6.6){\textcolor{blue}{\footnotesize{$\pi$}}}
\put(16.2,21.0){\textcolor{blue}{\footnotesize{$b_{1}$}}}
\put(23.6,34.4){\textcolor{blue}{\footnotesize{$\pi_{2}$}}}
\put(6.6,2.){\textcolor{red}{\line(0,1){3.6}}}
\put(13.5,2.){\textcolor{red}{\line(0,1){3.6}}}
\put(5.2,-.5){\small \color{red}$s_{0}$}
\put(12.2,-.5){\small \color{red}$s_{1}$}
\end{overpic}
%\end{figure}
%\end{center}
  \end{minipage}
  \vspace{-.2cm}
\begin{figure}[htb]
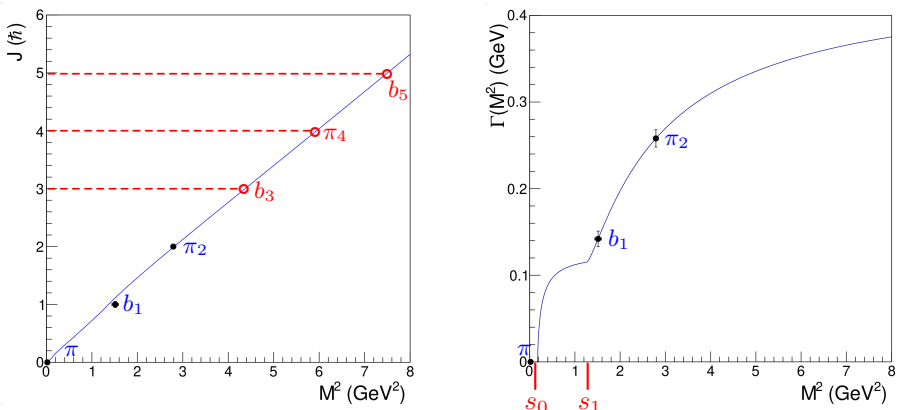

  \caption{Real part $(\pi,b)$-trajectory on the left, width function
  $\Gamma$ on the right.}
\label{Fig:pi_traj}
\end{figure}

\vspace{-.1cm}
In Fig. \ref{Fig:pi_traj} on the left, the three data points of the $\pi,b_{1}$
and $\pi_{2}$ states are shown by black points, and the non-linear fit by the
blue line. On the right, the widths of the $\pi,b_{1}$ and $\pi_{2}$ states are
shown by black points, and the fitted width function $\Gamma$ by the blue line.
The thresholds $s_{0}$ and $s_{1}$ used in the fit of Eq. \ref{eq:imag} are
shown in red. The thresholds \mbox{$s_{0}$=0.176 GeV$^{2}$ and}
\mbox{$s_{1}$=1.27 GeV$^{2}$} are defined by the decays
$\pi_{2}\!\rightarrow\!3\pi$ and $b_{1}\!\rightarrow\!K\bar{K}\pi$,
respectively. This fit of the ($\pi,b$)-trajectory predicts a $b_{3}$ state
with mass of 2090 MeV and width of 321 MeV,
a $\pi_{4}$ state with mass of 2437 MeV and width of 352 MeV,
and a $b_{5}$ state with mass of 2738 MeV and width of 371 MeV.

\section{The final-state resonance mass distribution}

The ($\pi,b$)-trajectory consists of $\pi$ and $b$-resonances with quantum
numbers (P,C)=($-,+$), and (P,C)=($+,-$), respectively. The final state
shown in Fig. \ref{Fig:mult_red} can hence contain two $\pi$-resonances,
or two $b$-resonances.

 \begin{minipage}[t]{.94\textwidth}
%%    \begin{center}
%  \begin{figure}
\vspace{-.2cm}
    \begin{overpic}[width=.44\textwidth]{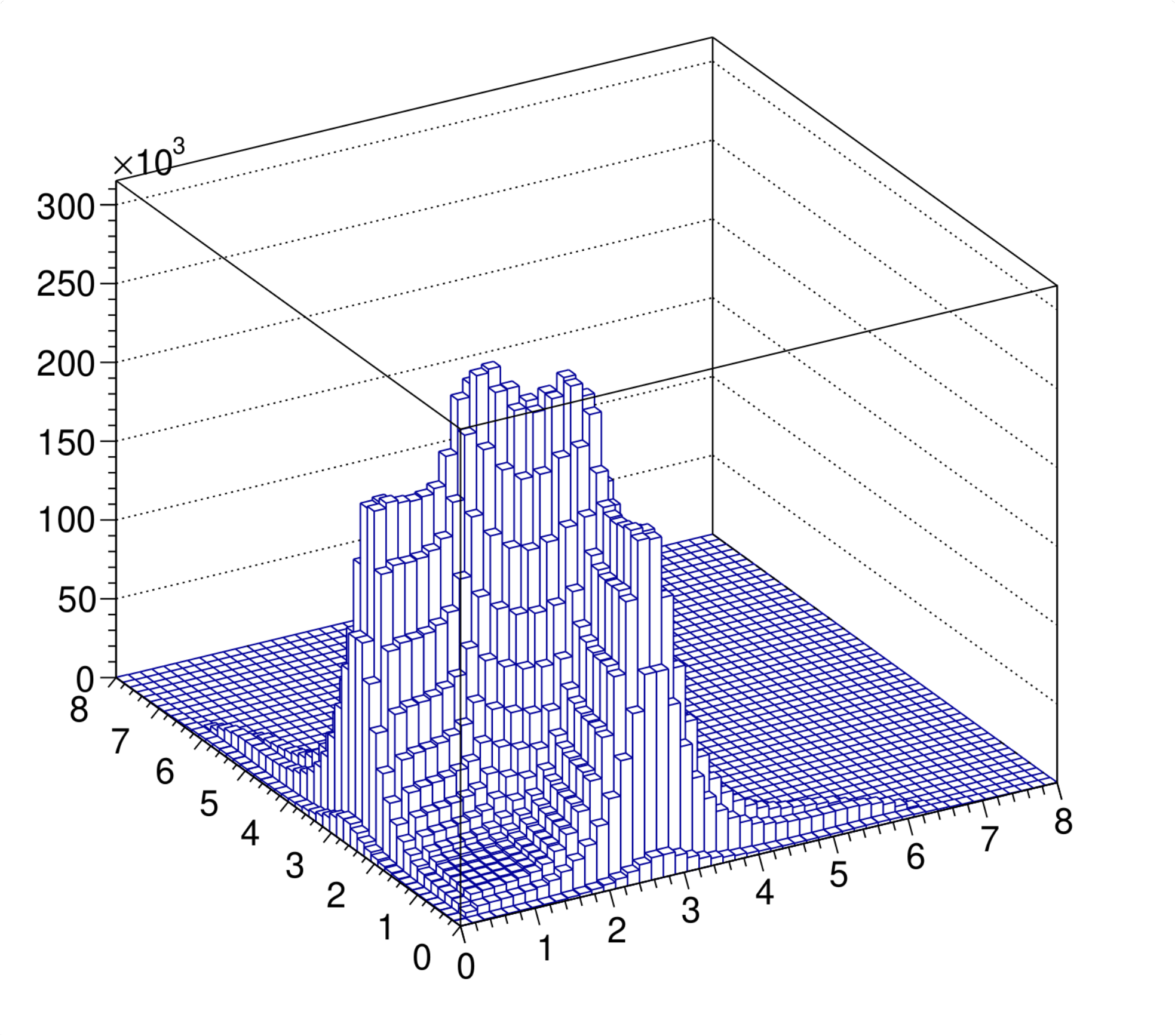}
      \put(16.,-.4){\tiny PC=(-,+)}
      \put(46.,4.){\rotatebox{16}{\tiny  $M^{2}_{1}(\tilde{S}^{\pi}_{1})$}}
      \put(2.,11.){\rotatebox{-30}{\tiny $M^{2}_{2}(\tilde{S}^{\pi}_{2})$}}
      \put(-3.,28.){\rotatebox{90}{\small arb. units}}
    \end{overpic}
    \hspace{0.1cm}
\begin{overpic}[width=.45\textwidth]{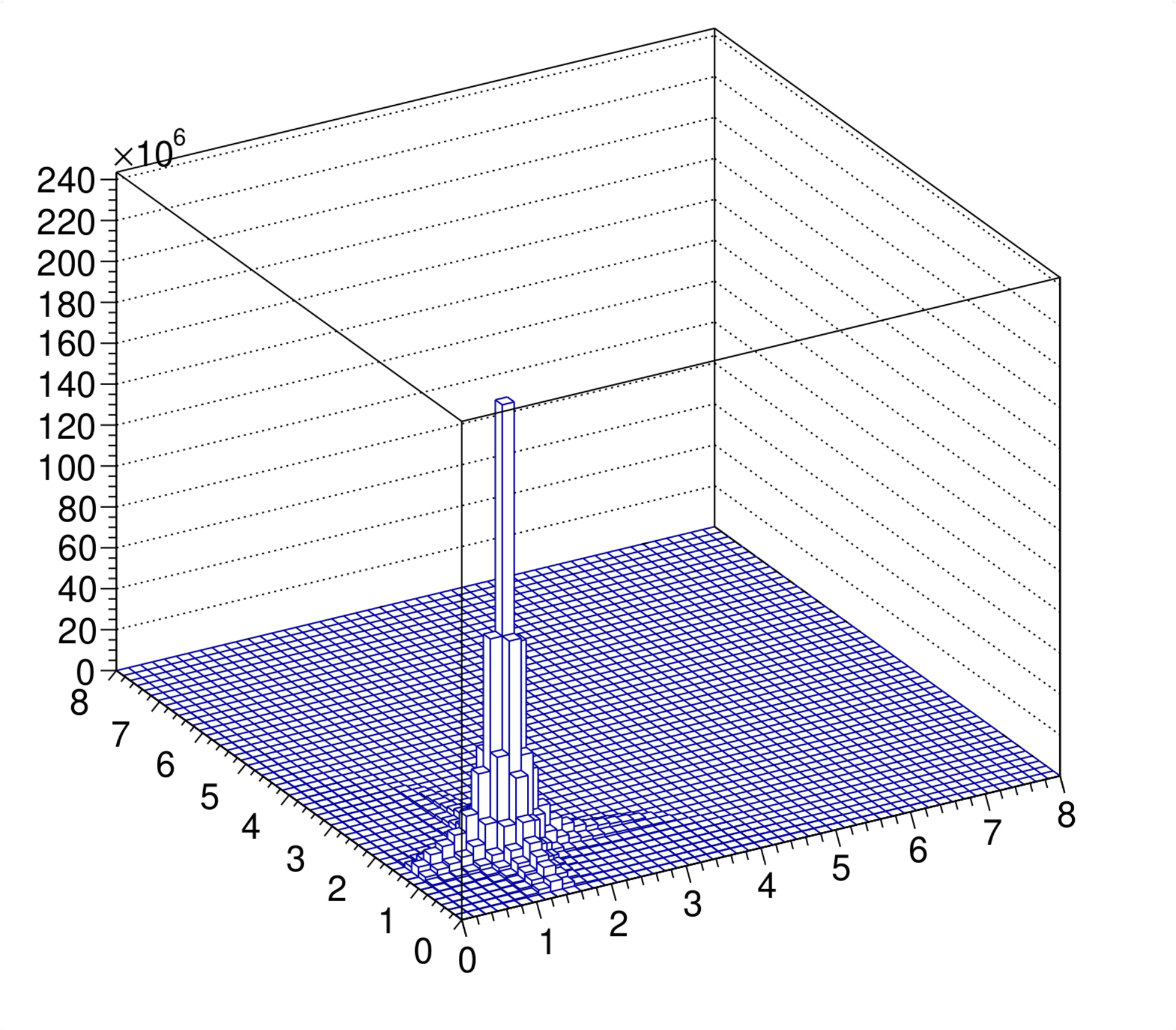}
\put(16.,-.4){\tiny PC=(+,-)}
      \put(46.,4.){\rotatebox{16}{\tiny $M^{2}_{1}(\tilde{S}^{b}_{1})$}}
      \put(2.,11.){\rotatebox{-30}{\tiny $M^{2}_{2}(\tilde{S}^{b}_{2})$}}
      \put(-2.,28.){\rotatebox{90}{\small arb. units}}
\end{overpic}
%\end{figure}
%\end{center}
\end{minipage}
\vspace{-.2cm}
\begin{figure}[htb]
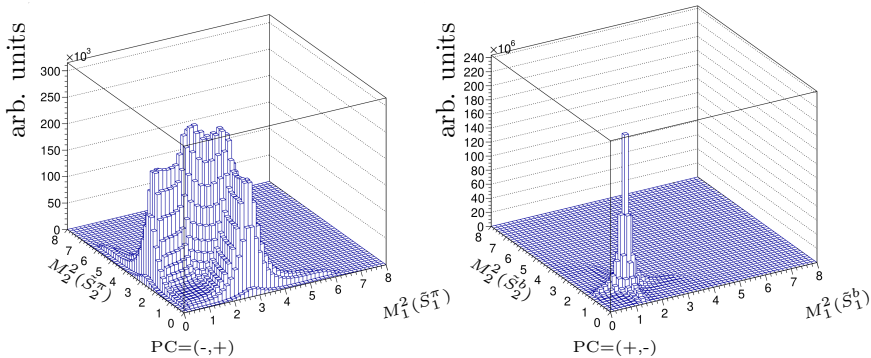

  \caption{Two-dimensional mass distribution of the final-state resonances.}
\label{Fig:2d_mass}
\end{figure}

\vspace{-.2cm}
In Fig. \ref{Fig:2d_mass}, the two-dimensional distribution of squared masses
is shown for $\tilde{s}$ = 9 GeV$^{2}$ for the case of two $\pi$-resonances
on the left, and the corresponding distribution for two $b$-resonances
on the right. Here, $\tilde{s}$ denotes the center-of-mass energy of the
two initial-state Pomerons as shown in Fig. \ref{Fig:mult_red}.

\vspace{-.18cm}
\section{Acknowledgements}

This work is supported by the German Federal Ministry of Education and
Research under reference 05P21VHCA1.
An EMMI visiting Professorship at the University of Heidelberg is
gratefully acknowledged by L.J.

\vspace{-.28cm}
\bibliographystyle{unsrt}
\bibliography{schicker_diff2022}

\end{document}